\documentclass[runningheads]{llncs}
\usepackage{graphicx}
\usepackage{tabularx}
\usepackage[utf8]{inputenc}
\usepackage{cite}
\usepackage{tcolorbox}
\usepackage{todonotes}
\usepackage{fancyhdr}
\usepackage{url}

\usepackage{booktabs}
\usepackage{array}
\usepackage{colortbl}
\usepackage{multirow}
\usepackage{longtable}

\newcolumntype{v}[1]{>{\raggedright \hspace {0pt}}p{#1}}
\newcolumntype{G}[1]{>{\columncolor{gray90}}#1}

\definecolor{Gray}{gray}{0.8}
\definecolor{gray25}{gray}{0.25}
\definecolor{gray50}{gray}{0.50}
\definecolor{gray75}{gray}{0.75}
\definecolor{gray90}{gray}{0.9}

\newcommand{\grayrow}{\rowcolor{gray90}}
\newcommand{\interviewquote}[2]{\begin{quote}
\footnotesize{\emph{``#1'' }} --- \footnotesize{#2}
\end{quote}}

\usepackage{booktabs}
\usepackage{array}
\usepackage{colortbl}

\newcommand{\etal}{\textit{et al.}}

\begin{document}
\title{
Agile Islands in a Waterfall Environment: Requirements Engineering Challenges and Strategies in Automotive
}
\titlerunning{Agile Islands in Waterfall}

\author{Rashidah Kasauli\inst{1,2}
\and
Eric Knauss \inst{1}
\and
Joyce Nakatumba-Nabende\inst{1,2}
\and
Benjamin Kanagwa\inst{2}
}
\authorrunning{R. Kasauli et al.}
\institute{Chalmers$\mid$University of Gothenburg, Sweden\\ \email{\{rashida, nabende\}@chalmers.se}, eric.knauss@cse.gu.se \and
Dept. of Networks\\
Makerere University, Uganda\\
\email{\{rnamisanvu, jnakatumba, bkanagwa\}@cis.mak.ac.ug}
}
\maketitle              
\begin{abstract}
{\bf [Context \& motivation]} Driven by the need for faster time-to-market and {reduced development lead-time}, large-scale systems engineering companies are adopting agile methods in their organizations.  
{This \emph{agile transformation} is challenging and} it is common that adoption starts bottom-up with {agile} software teams 
{within the context of} {traditional} company structures. 
{\bf [Question/Problem]} This creates the challenge of agile teams working {within} a {document-centric and plan-driven} (or waterfall) environment. 
While it may be desirable to take the best of both worlds, it is not clear how that can be achieved especially with respect to managing requirements in large-scale systems.
{\bf [Principal ideas/Results]} This paper presents an exploratory case study at an automotive company, focusing on two departments of a large-scale systems company that is in the process of company-wide agile adoption.
{\bf [Contribution]} We present challenges {related to requirements engineering that }agile teams {face} while working {within} a {larger} plan-driven {context} and propose {potential} strategies to mitigate the challenges. 
Challenges relate to, e.g., development teams not being aware of the high-level requirement and dealing with flexibility of writing user stories. We found that strategies for overcoming most of these challenges are still lacking and thus call for more research.

\end{abstract}
%
%
%
\section{Introduction}

In order to meet market demands and increasing competition, large-scale systems engineering companies are adopting agile methods 
~\cite{pernstal12,van2013agile,bjarnason2011case}.
{To leverage the full benefits of agility with respect to time-to-market and flexibility, such agile transformation must in the long run affect the overall systems engineering process and organization.}
While software teams may have {quickly} adopted agile methods and are working with them, there is usually slow company-wide adoption of agile methods, mostly attributed to the skepticism they received \cite{lindvall2004agile}. 
Since teams have to work with the traditional structures within the companies, this ends up creating `\textit{agile islands in a waterfall}' \cite{kasauli2017} -- a coexistence of both agile and traditional development approaches.  

However, without a company-wide adoption structure, different development teams adopt different practices \cite{wohlrab2018problem}. 
{Further, in the systems engineering context, software development is only one of several domains and must be synchronized with development of hardware and mechanics.} 
With teams contributing to the same product, it becomes challenging to deal with different speeds and quality measures from the different {ways of working}, especially with respect to Requirements Engineering (RE). 
{While there is a growing body of literature on agile transformations} \cite{almeida2017challenges,bannink2014challenges,dikert2016challenges}{, only a few studies address the coexistence of agile and traditional methods} \cite{van2013agile,kusters2017agile,theocharis2015water,kuhrmann2017hybrid}. 
{ We are not aware of any empirical work that explores such coexistence from the development teams' perspective{, especially in relation to requirements}.}
This study {fills} that gap by first understanding and documenting the challenges and lessons learned based on a multi-case study situated in the {automotive domain}. 
{This domain is particularly interesting, since requirements traditionally play an important role in automotive systems, yet recent trends around electrification, advanced driver assistance systems (ADAS) and autonomous drive (AD) as well as a market that moves towards continuous deployment of software functions render it necessary to change established engineering practices.}
Through an exploratory case study with two departments of an automotive company, based on a total of 18 interviews and one focus group, we explore the following research questions from a RE perspective:

\newcommand{\researchquestionone}{What are the perceived challenges {related to RE} when combining plan-driven and agile paradigms in large-scale systems engineering?}
\newcommand{\researchquestiontwo}{What mitigation strategies exist when using agile development in a traditional setting?}

\noindent \emph{\textbf{RQ1:} \researchquestionone}

\noindent \emph{\textbf{RQ2:} \researchquestiontwo}


In answering our research questions, {we provide important empirical data about the interplay of agile software development teams operating in a larger, plan-driven system development context from a requirements engineering perspective.} 
We present specific challenges with their proposed mitigation strategies. 
We believe that this study will benefit both researchers in the RE field and practitioners working in companies with similar environments.  

\section{Related Work}\label{related_work}
Traditional methods like waterfall and V-model have formed the foundation for systems development for decades. 
These methods follow a sequential execution of processes with predefined phases, extensive requirements design and documentation \cite{van2013agile,kusters2017agile}. 
In  traditional methods,  RE  includes  a  set  of  well  defined  preliminary phases dealing with analysis, planning, and documentation \cite{bucaioni2018alignment}.
Meaning that requirements are supposed to be complete in the preliminary stage.
However, faced with evolving market demands and fast changing requirements, systems development companies are adopting agile methods \cite{kasauli2017,theocharis2015water,kuhrmann2017hybrid} as the traditional methods are not flexible and are generally slow due to their rigorous nature.   

Agile methods encourage flexible and light-weight software development with short iterations \cite{Meyer2014}. 
Being flexible, agile methods can help where traditional methods fail \cite{kuusinen2016strategies}. 
{In agile development, RE is an iterative and continuous process that is carried out in every step of development. Also, requirements are only partially known and evolve rapidly during development.}
%
%
Previous studies on agile and traditional development have compared the characteristics of agile development with those of traditional methods\cite{boehm2005management,sillitti2005requirements,carson2013} while describing the way different organizations are adopting to these methods \cite{kuusinen2016strategies,theocharis2015water}. There are few studies concerning the coexistence of agile methodologies with traditional approaches \cite{van2013agile,kusters2017agile,theocharis2015water,kuhrmann2017hybrid,kuhrmann2018hybrid}.

Theocharis \etal \cite{theocharis2015water} study the general process use over time to investigate how traditional and agile methods are used. 
Is there coexistence or do agile methods accelerate the traditional processes' extinction? This was done by applying instruments of systematic literature review process.
They find indication to mixed application of traditional and agile methods and thus conclude that hybrid approaches that include traditional and agile approaches shape today’s “standard process ecosystem”. 
The results of this exploratory study show that there is less material published on the combination of software development approaches than expected. 
This is supported by Kuhrmann \etal \cite{kuhrmann2017hybrid,kuhrmann2018hybrid} who present results from 69 study participants in a survey on hybrid software development approaches. 
They found that companies combine different development approaches regardless of the industry sector and size. These findings go beyond the adoption problem to having a state of working with both methods in coexistence. 
This confirms the coexistence of both methodologies.

Based on a Grounded Theory research involving 21 agile practitioners from two large enterprise organizations in the Netherlands, Waardenburg and van Vliet \cite{van2013agile} present 
the challenges of using agile methods in traditional enterprise environments. 
They organized the challenges under two factors: increased landscape complexity and lack of business involvement, for which they identify successful mitigation strategies. These mitigation strategies concern the communication between the agile and traditional part of the organization, and the timing of that communication, an observation also made by Eliasson \etal \cite{eliasson2015}. Kuusinen \etal \cite{kuusinen2016strategies} investigate practitioners' mitigation strategies related to the challenge of doing Agile in a non-Agile environment. Through an on-line survey and a workshop, they provide strategies from both an organizational and change perspective. The strategies include: (a) ensuring managers understand and buy-in to Agile and (b) creating an organizational culture that fosters agility as the biggest themes of the study. Whereas both these studies give the organizational view, we provide the developers' perspective.

Starting from a literature search, Kusters \etal \cite{kusters2017agile} derive risks and problems at the interface of agile and traditional development approaches in hybrid organizations which have an impact on coordination and cooperation. They discover 28 issues which reduce to 22 after validation through a case study at a large financial institute in the Netherlands. They had six classifications of challenges including challenges relating to development and testing. We explore challenges in this classification, particularly development. 

In summary, the body of knowledge on the coexistence of agile and traditional methods is growing. However, to the best of our knowledge, there is a lack of empirical studies that explore developers' experiences in this context and less from an RE perspective. We believe this study will help inform a better approach to the coexistence dilemma.


\section{Research Method}\label{research_design}
{In this study, we} explore development challenges {related to RE} {caused by the mixture of} agile and plan-driven methods{.}
{We rely on} the case study method \cite{runeson2012case}, which is considered most appropriate when studying {a complex social} contemporary phenomenon{, for which deep understanding of the context is critical} \cite{yin2009case}. 
{Specifically, we investigate} the case of an automotive company and using two of its departments as units of analysis{.
We} collected data {through 18} semi-structured interviews and used thematic coding for analysis. 
This section details our data collection and data analysis steps and also discusses the validity threats. 

\textbf{Data Collection}\label{sec:data}
The study was conducted at Company X, a large automotive manufacturer (OEM) distributed across several countries. 
The study was done at two different departments (A and B) in this company, which 
are {among} those that pioneered the company{'s increased} move to in-house software development{ with the aim to further increase flexibility and to decrease the lead-time for late changes}. 

The first series of 11 interviews with Department A was conducted as part of a master thesis\footnote{Though supportive, the student was not able to be part of this continuation.}\cite{gopakumar2016challenges} and motivated by those results and our parallel research on challenges of RE for large-scale agile \cite{kasauli2017}, we were encouraged by Company X to replicate a similar investigation with a different department.
The first two authors re-visited the data previously collected and followed up with a focus group at Department A and preparation meetings throughout Company X and at Department B, which was chosen through maximum variation strategy  \cite{palinkas2015purposeful}.
The first author of this paper then conducted 7 interviews at Department B.



\urldef\igurl\url{http://www.cse.chalmers.se/~knauss/2020-AgileIslands}
{Throughout the different phases of this research (including the master thesis), the authors collaborated in} selecting participants and defining the interview guide based on consultations with our main contacts in both departments.
Participants were selected from two areas: a) the agile software teams and b) the plan-driven system level roles. 
Both areas think very differently about development: the agile software teams did not necessarily like to talk about requirements, while the system level roles did not feel that a discussion of agile artifacts would provide value. 
This was mainly a problem at Department A, where we choose to avoid using the word ``requirements'' in our interview guide for agile software teams where necessary. 
In Department B, this aspect was deemed less critical, but since the way of working differed significantly, we had to adjust the interview guide\footnote{Interview guides: \igurl} 
again for their context -- allowing us to follow up on findings from Department A. 
{We discuss each department in more detail in section \ref{case_study}.}



{All} interview guides contained questions which sought for information on how requirements flow from one role to another within the development team. 
Each interview started with explaining the purpose of the interview and  interview time ranged from 45 min to 1 hour. 
All interviews were voice-recorded and notes were taken during the interviews.
The roles {and responsibilities of }interview{ees} for each of the departments are presented in Table \ref{tab:sources}.


\begin{table}[ht]
\centering
\caption{Roles and responsibilities of participants in the study.}
\label{tab:sources}
\begin{tabular}
{v{0.3\textwidth}v{0.6\textwidth}|c|c}
\toprule
\multirow{2}{*}{Role} & \multirow{2}{*}{Responsibilities} & \multicolumn{2}{c}{Dept} \tabularnewline
\cline{3-4}
& & A & B
\tabularnewline
\midrule
\grayrow Function Owner (FO)& Owner of one or several car functions and produces the high-level requirement(s) for the function. & 2 & 2
\tabularnewline
Function Realisator (FRR)& Breaks down the FO requirement to a slightly more detailed description of how the function should be realized and distributes to different sub-systems.&&1 
\tabularnewline
\grayrow System Designer (SDE) & Describes the high-level requirement and design of the sub-system. &1&1 
\tabularnewline
 System Responsible (SR)& Produce a detailed design of the software and hardware and also write detailed requirements (SRS).&1&1 
\tabularnewline
\midrule
\grayrow Software Quality Engineer (SQE) & Comes in when something goes wrong to ensure that correct design is followed & 1 &
\tabularnewline
\midrule
 Testers (SFT)& Depending on what level the tester is, they are responsible for writing a test \& verification plan that describe the verification steps to cover the requirement, and then perform verification/validation and report the results.&2& 
\tabularnewline
\grayrow Scrum Master (SM) & ``Responsible for ensuring the team lives agile values and principles and follows the processes and practices that the team agreed they would use''  \cite{AgileAl2019} .
&1&
\tabularnewline
 Product Owner (PO)& Manages and prioritises backlog.&1&
\tabularnewline
\grayrow Developer (Dev) & Create Software solution.&2&2 
\tabularnewline
\midrule
\end{tabular}
\end{table}

\textbf{Data Analysis}
The interviews were transcribed and thematic coding method  \cite{gibbs2008analysing} was used to identify, analyze and report patterns within the data. 
Transcriptions were used to identify, name and categorize phrases and words in order to develop the initial codes. 
Initial coding enabled the generation of inductive codes which were later grouped into themes. 
The themes were {iteratively and collaborativly} refined {by all authors} and named according to the aspects addressed in the interview guide. 
We thus agreed on the following general themes; requirements engineering process (which included the teams involved and roles), challenges faced, and solutions proposed.

\textbf{Threats to Validity}
{Three researchers worked iteratively with codes and their grouping into themes, during which potential misinterpretations were raised and discussed.}
The initial results were discussed in a focus group at Department A that included four key roles already involved in the study and {two} more people responsible for requirements management. 
Results were presented and discussed {to check whether we misinterpreted any data and to provide an opportunity to raise any challenges we may have missed.} 
{We then discussed} strategies to overcome {the challenges} {with} the participants. 
For Department B, we shared the findings through email and when clarifications were needed, calls were made. 
To increase internal validity, we used data triangulation between the case departments. 
We study {two departments within }one case {company in a qualitative exploratory case study}, thus {sacrificing} generalization of our findings to other domains {in favour of in depth-knowledge in the specific case}.
{We believe that this will enable future research in the area.} 
{We believe by discussing } the results with company contacts in both departments and {by comparing them} to existing problems in literature{, we mitigate threats to reliability of our findings}. 
\section{The Case Study}\label{case_study} 


Company X follows a hierarchical structure and employees are departmentalized, following a chain of command. 
{Software is partly developed in-house and partly purchased from suppliers.}
Recently, however, software development is increasingly moved in-house as the company endeavors to shift from the traditional methods of working to being more agile development oriented in all working departments.


{In earlier work, we found system engineering organizations differ in the scope as they introduce agile methods} \cite{kasauli2017}.
{While few companies have achieved some level of agility for the full organization, at the time of this research, many had only introduced agile methods on the level of agile software development teams, while keeping a plan-driven approach for the overall systems engineering process.}
{Our case company is no exception, yet both departments covered in our study differ in how they manage the interplay of plan-driven systems engineering and agile software development.}
%
%
%
{While both departments have significant experience with agile software development to support their business goal of increased flexibility, they differ in the type of software to be developed.


{\bf Department A}  is responsible for development of {(often safety-critical)} functions of the {overall} vehicle's system. 
Their development focuses on algorithms that connect large amounts of vehicle data with cloud intelligence, and agility helps to increase learning about how these algorithms behave in a realistic context early on}, comparable to a system that we previously discussed \cite{KARI2015}.
While suppliers for hardware, for example chip-sets exist, much of the software is done in-house to increase flexibility, decrease lead-time of new features, and protect innovations and intellectual property. 

{\bf Department B}, {in contrast, is responsible for the development of a central platform that is the foundation for the work of several other departments. 
Here, agility is employed to prioritize change requests from many stakeholders and in this way maximize the flexibility of overall system development.}
Thus, Department B has potentially huge impact on flexibility and lead-time for the whole system. 

In order to give context to our findings, we extracted an overview of the current RE process at both departments, which we describe in terms of the roles and responsibilities (Tab. \ref{tab:sources} and Fig. \ref{fig:coexist}) 

\subsection{Roles in the Departments}
The requirements process in the two departments is executed through different roles {in a strict hierarchy,} with requirements going back and forth between different levels, for clarification and update where necessary. 
There is a lot of communication, both formal and informal in both cases. 
The roles can be grouped in two categories; those that make up the hierarchy and roles that constitute the software development team. 

The hierarchy levels start with the Function Owner (FO) who receives the customer function and creates a high-level document which the Function Realizer (FRR) uses to distribute requirements to the different subsystems. Depending on how scalable the requirements are, the FRR 
could work with the same requirements or sometimes break them down before assigning them to the subsystems. 
At the subsystem level, the System Responsible (SR) breaks the requirements into software (SWRS) and hardware requirements. The software requirements are sent to the in-house team while the hardware requirements are sent to the respective suppliers. At the same level, the System Design Engineer (SDE) designs the technical solution, i.e, determine how to implement the requirement and how to allocate it on different components which the in-house development team implements. The SR and SDE work together and have their own iterations as they write the requirements and respective technical solutions. This concludes the hierarchical levels.

The roles in the lower part in Table \ref{tab:sources} relate to software development team as depicted at the bottom of Figure \ref{fig:coexist}. 
Here the departments take different approaches to agility. 
Department A, has a Product Owner (PO) who is responsible for managing and prioritizing the backlog and then puts the requirements in Jira for implementation by the developers. 
In case of any change requests, the development team communicates to the PO who discusses with the FO to get the requirements changed. So the plan-driven  requirements do not seem to {directly} interact with the agile process of development.
In Department B, the SDE is the one that creates an \textit{issue} in Jira which the developers use to implement the requirement. The software issue contains the `user story' and the software requirement from SWRS. In case of any disagreements, the developers have to go through the SDE who then discusses with the FO, if necessary. 

\subsection{Requirements Model in the Departments}
The FO produces a requirement document {to specify the} function and shares it with the FRR. 
There is handshaking of the requirement between FO and FRR through back and forth communication to verify the requirement before it is broken down and sent to the next level. 
Due to the iterative process, the requirement document keeps changing and the company contacts preferred to term it the \textit{requirements model}.

\begin{figure}[htb]
\centering
\includegraphics[width=0.6\textwidth]{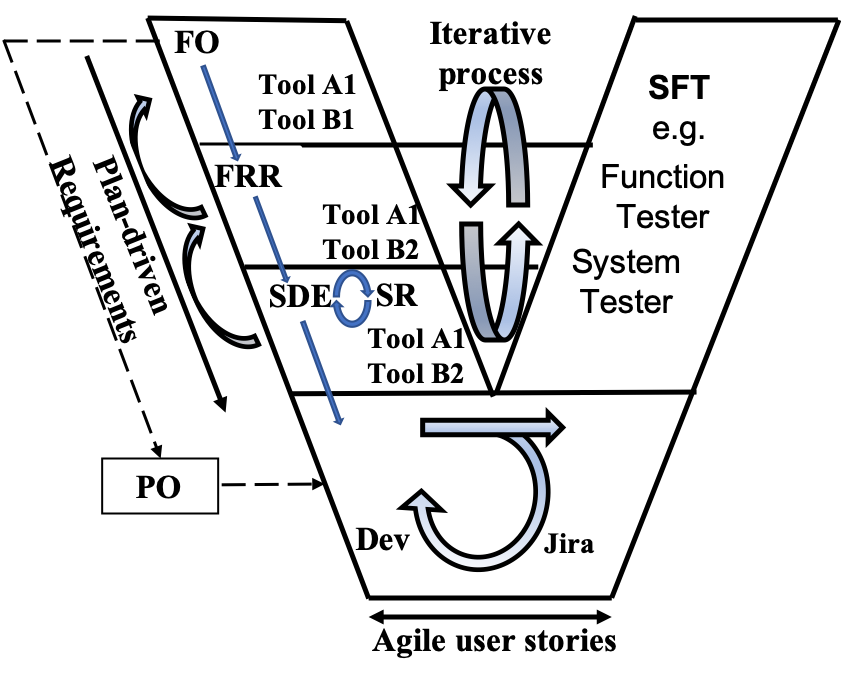}
\caption{Requirements and user-story coexistence model.}
\label{fig:coexist}
\end{figure}

Figure \ref{fig:coexist} shows {what our contacts call} the \textit{Narrow}-V model, the hierarchical model followed by both departments. 
{It keeps the abstraction levels from the standard V-model \cite{balaji2012waterfall} and has three general layers of requirements, but emphasizes iterative work on system level.} 
On the left hand side, the requirements are formed and broken down at the different levels while the right hand side shows testers that write corresponding test cases for the requirements. 


The \textit{Narrow}-V model is supported by different tools in the two departments. Department A uses Tool A1 for all the plan-driven levels while Department B uses two different tools; one for FO level (Tool B1) and another for FRR and SDE levels (Tool B2). For the agile part, the development part, both departments use the Jira tool to support their implementation. 
Anyone with access to Jira can write a requirement{ or }user story to the developers in either case but the team only works on those approved by their immediate superiors, the PO for Department A and SDE for Department B. 
\section{Challenges and Strategies}\label{challenges}
{This section describes the challenges {and potential solutions} identified {through interviews and discussions in a focus group}. 
These challenges are presented in three parts: In \ref{sec:both} the challenges common to both departments are discussed while \ref{sec:unicA} and \ref{sec:unicB} discuss those challenges unique to department A and B respectively.}
For each challenge, we discuss the corresponding potential strategies.

\subsection{Challenges in Departments A and B} \label{sec:both}
 {Three challenges were mentioned by more than 50\% of the interviewees in both departments and these are described here.}

\textbf{C.1: Development team not aware of the requirement model.} 
An issue raised by developers in both departments is the lack of awareness about the FO's requirements. 
In Department A, only the PO is both in direct contact with the development team and has knowledge about the FO's requirements. 
Even though the developers have access to Tool A1 where the requirements are stored, they are rarely accessing them.
This is partly due to them using a different tool in their daily work, which relates to a challenge of developers in Department B, who get their requirements from the SDE and have no access to the FO's requirements in Tool B1.
One developer mentioned not getting to know what the legal requirement is, which only the FO could know from the requirements they have written. 
According to developer at Department B: 
\interviewquote{...to know that if its a bug for example, is this something breaking the law or could I wait to implement this bug fix. Could we send this software with this bug fix or is it against the law I don't know.}{Dev-B}

It is however also a challenge that the requirements, as they are currently specified, do not appear to fit the agile way of working and do not easily translate into concrete development tasks on the backlog. 
The differences in how FO and the development team think about requirements appear hard to bridge at the time of this case study.
With the pending proposals to include (system-level) testers in the development team, it is anticipated that this challenge will grow.

\textit{Proposed Solution:} According to the interviewees, system-level testers generally have a stronger awareness of high-level requirements than developers. 
Thus, one solution proposal is to have more system-level testers as part of the agile development team.
However, this fusion can only succeed if the agile team members are open to discuss requirements and to increase their awareness.
Also, it may  be hard to find enough experienced system-level testers for each team.  
Thus, interviewees felt a need of forcing the agile teams to create formal test report as part of their work, where test results are related to high-level requirements, forcing them to read the requirements model. 
A good compromise, would be not to do that every sprint but only for important releases of the software component. 
Obviously, this demands for teams to get access to the requirement model, which can be an organizational challenge.



\textbf{C.2: FO over exposed to change requests.} \label{sec:FO}
{In a traditional hierarchical and plan-driven approach, FOs discuss requirements mainly with the FRR and few change requests are made, mostly top-down and very rarely originating bottom-up from the development team.}
With agile development, however, comes a higher commitment to respond to change and development teams need to take more responsibility in clarifying and refining requirements.
Proposed solutions to C.1 also include increased awareness and access to requirements for the development teams. 
This however creates a challenge to the FOs with change requests and opinions coming more frequently and increasingly from the development teams.

In Department A, even though hierarchies do exist, everyone seems to have the right to make a change proposal to the FO. As one FO comments:

\interviewquote{I would say it is a good thing that many people read the requirements. But then it also means there is going to be more opinions, comments and also more work.}{FO-A}

Both FOs in Department A however agree that there is a danger to become the bottleneck in this process.

In Department B, the hierarchy is stricter. 
For instance, developers  do not get to propose changes directly to the FO, also since they do not even have access to his requirement (see C.1).
Still, the FO at Department B  receives many questions 
{about the function}
{ and there is a tendency to involve the FO more in FRR and SDE level work to help making decisions about changes.} 
The exposure to change requests may not allow time to perform other activities and FOs find themselves at risk to become bottlenecks of the process. 

\textit{Proposed Solution:} {All} Interviewees {agreed that in an agile way of working} requirements {should be updated \emph{bottom-up} based on learning} from the Sprint. 
{A high number of change requests must be embraced by an organization that aims to fit agility into their system development.}
This could also help to improve 
requirements with arbitrary performance goals.  
{Yet, there is no clear, non-trivial solution (like adding more resources) to remove this bottleneck.} A few of the interviewees, hinted on having the team more involved in updating or changing the high-level requirement. 

\textbf{C.3: {`Suffering'} Requirements traceability.}
{The traceability between requirements artifacts used in the two departments is suffering, with differing levels of impact for each department.}
In Department A, the PO {writes users stories from FO requirements and the team decides how to implement them.}
The team members can also decide to breakdown the stories to more granular levels and put to the backlog for discussion and later prioritisation.
Clearly, not all user stories need to be traced, however, it is not clear which user stories should be traced to requirements. 
This can lead to additional effort, when traceability must be ensured, 
 since important information is missing and must be reconstructed.
When asked to comment on traceability, one developer responded that: 
\interviewquote{I don’t think traceability is not required or something like that. It’s just that my focus hasn’t been on documenting the function. I just focus on doing implementation and developing the function.}{Dev-A}
The respondents also mentioned that it is unclear whose responsibility it is to document that traceability. 

In Department B, 
the SDE creates a `\textit{software issue}' and all updates and changes affecting that requirement are tracked in that issue. 
{A Software Requirements Traceability Matrix (SWRTM) is used to capture the implementation status of requirements and to link them to unit tests, revisions of  requirements, and suggested but unaccepted changes.}
This document is also used by function testers. 

{Despite these efforts, traceability suffers from the quick pace in agile development, especially since refinements of requirements on development team level are managed in a separate tool. 
Thus, in both departments a lot of effort goes into maintaining high quality traces, slowing down the speed of managing changes and challenging the intended goals of transitioning to agile.}

\textit{Proposed Solution:} All participants agreed to make explicit to a user story whether it is relevant for tracing, but requisite that requirements model is understood. 
One way is to have a clear definition of `done' for a given user story, that is, a user story as `done' only when it has proper information about tracing, e.g. `does not require tracing'. 

\subsection{Challenges Unique to Department A} \label{sec:unicA}
\textbf{C.4: All can write user stories anytime.}
In agile development, 
user stories can typically be proposed by anyone in the development organization, although it is the PO's responsibility to prioritise them in the backlog. 
Interviewees noted that although a standard process is followed once a user story arrives into the backlog, there are no proper channels for screening user stories before they get into the backlog, as anybody is allowed to write user stories. This results in several unimportant or wasteful user stories creeping into the backlog which takes up a lot of time in discussing them.

In Department B, respondents agreed that all can write user stories if they have access to the system. 
However, since the structure is rather strict here, developers focus to implement based on tasks received from the SDE and are generally not concerned with new user stories. 
Once a new user story is written, the System Team (made up of FO, SDE and SR) discuss{es} to reconsider the requirement. 
{If needed, the FO effects the change and developers do not get involved in managing new user stories.}

\textit{Proposed Solution:} There were proposals to create proper channels for user story writing, where some authentication is required before writing, something that corresponds to what is done in Department B. 
{This might however conflict with attempts to empower the teams more to manage requirements.}


\textbf{C.5: Inconsistency between requirements model and user stories implementation.}
The interviewees from SFT stated that they do receive all the requirements from SDE, but they {do not} get information on whether the requirements were 
{not implemented or changed}. 
So, they work on {the} assumption that all the prior test cases still are valid. 
However, during testing, they discover that either many of the requirements have not been implemented or the requirements do not match the implementation. 
This mismatch was attributed to failure to update requirements. 
The user stories change a lot during implementation and yet the traceability is not clear (see C.3). 
Furthermore, for the new user stories which arise during implementation, the corresponding requirements have not been written. 
Since traceability is lacking, requirement updating does not happen{, resulting in unnecessary additional work to establish a defined state before release}. 

{In contrast, }for Department B, the traceability problem surfaces on a higher level and similar inconsistency is not so evident.

\textit{Proposed Solution:}
Making testers as part of the development team was one strategy to overcome the challenge of inconsistency. 
This however {entails also a cultural change and} does not alleviate the problem at higher level. 

\subsection{Challenges Unique to Department B}\label{sec:unicB}
\textbf{C.6: Complete picture and system thinking missing.}
Traditionally, the teams had the full requirements and {generally knew} how they relate to each other. 
In agile development where the focus is on features, teams lose focus of how their functions relate to other functions.  
On the other hand, many different departments are using the same signal in different ways. One interviewee noted that some subsystems listen in to the signal without requesting for it. This makes it hard to know what effects a change in one unit could have on another unit. 

\textit{Proposed Solution:} This seems to be a classical problem of development especially when iterative. All interviewees agreed that maybe having a complete picture may not be practical but recommended visualisation of how each and every function is related to each other.

\section{Discussion and Conclusion}
\begin{table}[tb]
    \centering
    \caption{Summary of Challenges and Proposed Strategies}
    \label{tab:results}
    \begin{tabular}{v{0.3\textwidth}v{0.36\textwidth}v{0.3\textwidth}}
\toprule
\multirow{2}{*}{Challenge} & \multicolumn{2}{c}{Strategies} \tabularnewline
\cline{2-3}
 & Interviewees & Literature
\tabularnewline
\midrule
\grayrow C.1 Development team unaware of requirement model & Have testers as part of agile team\\ Team creates formal test report for release & Cross-functional teams \cite{bjarnason2011case, liebel2018organisation} \tabularnewline
C.2 FO over exposed to change requests & (Empower team to) update requirements based on learning from sprint & Cross-functional teams for requirements update \cite{bjarnason2011case,knauss2018t} \tabularnewline
\grayrow C.3 Suffering requirements traceability & Explicitly define if user story must be traced\\
Force team to understand reqts. model & Increase understanding of process and roles \cite{liebel2018organisation}\tabularnewline
C.4 All can write user stories anytime & Create proper channels for writing user stories & Cross-functional teams update requirements and peer-review \cite{knauss2018t}\tabularnewline
\grayrow C.5 Inconsistency between requirements model and user stories implementation & Have testers as part of team & Bring testers closer to requirement owner \cite{uusitalo2008linking}\\iterative reqts 
management \cite{de2017challenges}\tabularnewline
 C.6 Complete picture and system thinking missing & Visualisation of how functions relate & \tabularnewline
\bottomrule
    \end{tabular}
\end{table}

Evidently, these two departments have adopted agile methods in different ways, 
and rely on different practices.
%
Department B is more strict on user story management while Department A 
gives more freedom and responsibility to agile teams.
%
%
Both departments embed agile teams in similar hierarchies defined from plan-driven systems engineering, but differences in the organizational interface affects the severity of requirements-related challenges faced.
%
Department A team has a bit more autonomy than in Department B since they can propose changes to FO, something unheard of at Department B.

Interviewees proposed strategies for the challenges which we now compare with those proposed in literature in Table \ref{tab:results}. In Table \ref{tab:results} we see that there are few studies with solutions that directly link to the challenges identified. The cross functional teams have been proposed in literature to handle challenges C.1 and C.2. For this the teams would have to comprise of the required expertise to handle the update issues to FO and also manage high-level requirements within the team. 
For C.3 where it is not clear whose {responsibility} traceability is, Liebel \etal \cite{liebel2018organisation} recommend increased understanding of processes and roles to address the challenge of 'unclear responsibilities and borders'. 

Though some potential solutions are recommended in literature, 
there is a lack of empirical investigation on their ability to mitigate requirements related challenges, as for example with respect to cross-functional teams. 

In conclusion, we present results of the coexistence of agile methods and traditional, plan-driven methods at two departments of an automotive company.
By studying the requirements flow 
with focus on the roles and structures used to get the requirements across, we were able to identify challenges brought about by such coexistence. 
In both departments, and basing on the challenges observed, this coexistence of agile and plan-driven approaches seems to limit the 
{efficiency of development}.
The departments have different 
ways of working and use different tools. 
Each team implements agile methods in their own way, which leads to some differences in the challenges we observed. 
Generally, all challenges relate to working with requirements updates when the team is using agile methods while the department structure is plan-driven. 
%
%
Strategies to mitigate these challenges range from having cross-functional teams, improving traceability through adequate tooling to improving process understanding and having well defined roles that may include empowering teams to manage requirements on their end.
Practitioners in similar settings can use our results to facilitate process improvement and to drive agile transformations.
Future research will have to show to which extent these or other solutions can mitigate the challenges and to get most out of the combination of agile and plan-driven approaches.

\section*{Acknowledgments}
We thank all participants in this study for their great support, deep discussions, and clarifications. Special thanks goes to Swathi Gopakumar for her help in data collection. This work was supported by Software Center {Project 27 on RE for Large-Scale Agile System Dev.} and the Sida/BRIGHT project 317 under the Makerere-Swedish bilateral research programme 2015-2020.

\bibliographystyle{splncs04}
\bibliography{refsq19}

\end{document}